\documentclass[11pt,a4paper]{article}
\usepackage[hyperref]{nlp4MusA}
\usepackage{times}
\usepackage{latexsym}
\usepackage{graphicx}
\usepackage{color}
\usepackage{amsmath}
\usepackage{booktabs}
\usepackage{makecell}
\usepackage{subfigure}
\usepackage{multirow}
\usepackage{url}
\usepackage{float}
\usepackage{url}

\aclfinalcopy 

\title{Symbolic Music Playing Techniques Generation as a Tagging Problem}

\author{Yifan Xie, Rongfeng Li \\
  Beijing University of Posts and Telecommunications, Beijing, China\\
  \texttt{\{yifan.xie, lirongfeng\}@bupt.edu.cn} \\}

\date{}

\begin{document}
\maketitle
\begin{abstract}
  
Music generation has always been a hot topic. When discussing symbolic music, melody or harmonies are usually seen as the only generating targets. But in fact, playing techniques are also quite an important part of the music. In this paper, we discuss the playing techniques generation problem by seeing it as a tagging problem. We propose a model that can use both the current data and external knowledge. Experiments were carried out by applying the proposed model in Chinese bamboo flute music, and results show that our method can make generated music more lively.

\end{abstract}

\section{Introduction}
Music generation has always been a hot topic. As early as the classical music period, Mozart used the method of rolling dice to automatically generate music. In recent years, most of the automatic music generation methods are related to deep learning, various kinds of model such as the encoder-decoder framework \cite{yang2017midinet}, generative adversarial networks (GAN) \cite{dong2017musegan}, variational autoencoders (VAE) \cite{hennig2017classifying}, long-short-term memory (LSTM) \cite{hadjeres2017deepbach} and recurrent Boltzmann machines (RBM) \cite{10.5555/3042573.3042813} are used widely. 

Unfortunately, previous studies about music generation hardly ever took playing techniques into account at the symbolic level. In fact, playing techniques are also quite an important part of the music. For example, In Chinese bamboo flute music, different music styles have different playing techniques, which is helpful to better show the features of different styles. Even in some kinds of music, the playing techniques are more important than the melody. For example, scores of Guqin, can have no stable tonality and no stable duration of the pitch but must have definite playing techniques recorded. 

In this paper, we discuss the symbolic music playing techniques generation problem. We solve this problem by seeing it as a tagging problem. There is much discussion about tagging problem in natural language processing. Some sequence tagging models like Conditional Random Fields (CRF) \cite{lafferty2001conditional}, Bidirectional LSTM (BiLSTM) \cite{graves2013hybrid} and BiLSTM with a CRF layer (BiLSTM-CRF) \cite{huang2015bidirectional} perform well in many tagging tasks. However, they are only purely data-driven learning. In fact, especially for music, human perception of them depends not only on learning data from the current scene but also on some more general knowledge from the past. For example, when you try to study a new instrument, you can not only benefit from some knowledge from the current instrument which you are learning but also can benefit from some music knowledge which you have learned in the past. Therefore, we propose a model that can use both current data and external knowledge.

Our proposed framework is composed of three parts. The first part is studying the current data. In this part, a general sequence tagging model (like CRF, BiLSTM, BiLSTM-CRF, and so on) is used. The second part is studying external knowledge. In this part, external knowledge is first constructed into logic rules, then a weight matrix that implies logic rules is generated using an algorithm. The third part is to combine the previous two parts through matrix operations. We evaluate our model using a Chinese bamboo flute music dataset, and the results show that our method can make generated music more lively.

To the best of our knowledge, we are the first to explore playing techniques generation at the symbolic level, another contribution is that we propose a playing techniques generation model that can use both current data and external knowledge.

\section{Task Description}
\begin{figure}[H]
	\centerline{
		\includegraphics[width=155pt]{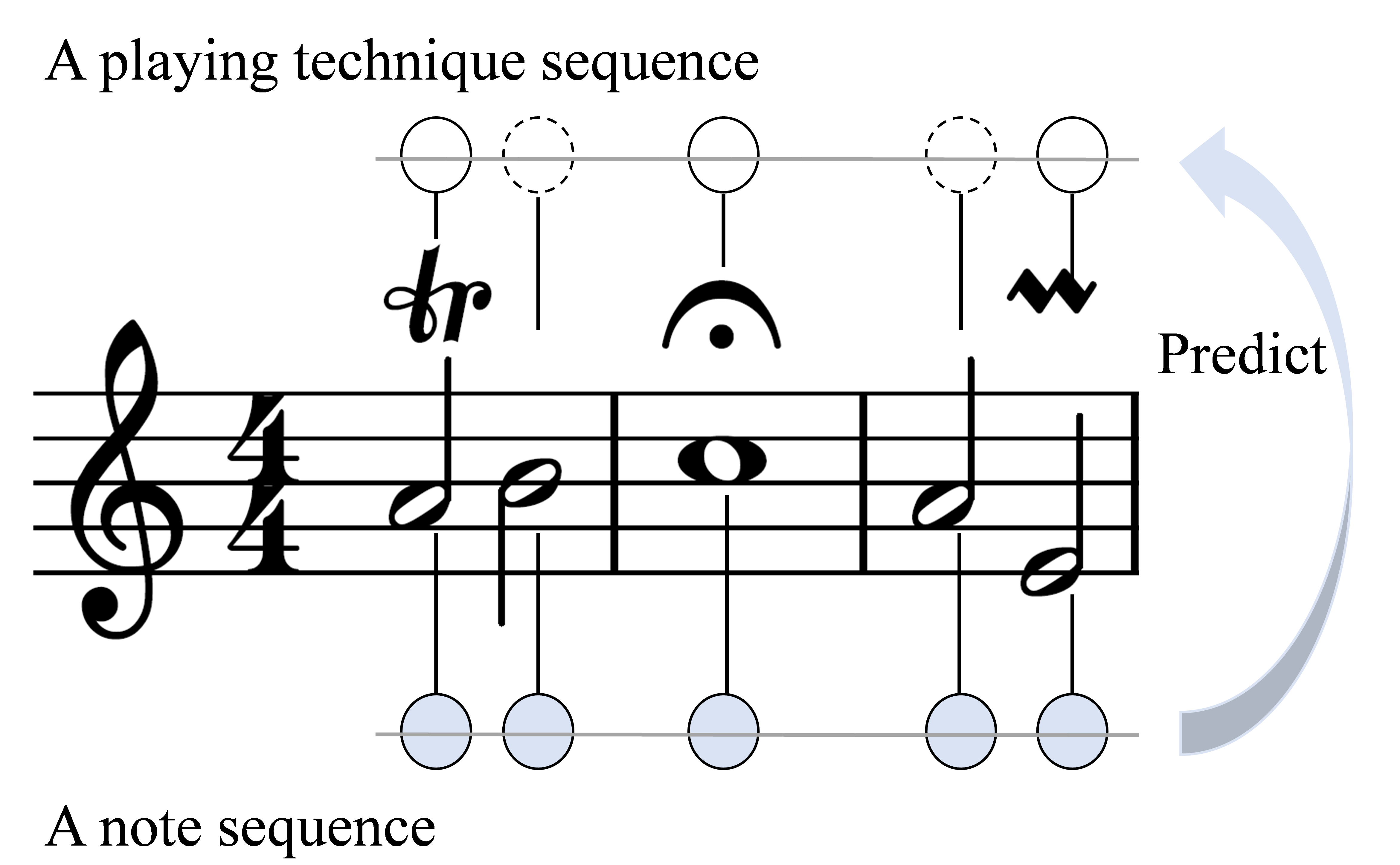}}
	{\caption{Task description}  \label{fig:task} }
\end{figure}

In this paper, we focus on monophonic music, but it can also be extended to polyphonic music easily. We see the playing techniques generation problem as a tagging problem, which consists of two processes. The first process is training a tagging model from a training dataset, the second process is applying the trained tagging model into a testing dataset to generate playing techniques. The goal of the tagging problem is that, given an observation sequence input, a tagging sequence (a state sequence) output can be predicted. In the playing techniques generation problem, a note sequence represents an observation sequence and a playing technique sequence represents a tagging sequence (see Figure \ref{fig:task}). Which is to say, playing techniques can be generated based on note sequences and a trained tagging model.

\section{Data Representation}

A monophonic melody can be seen as a note sequence. In this paper, each note is composed of the following features:

\begin{itemize}
	\item \textbf{Pitch}: Chromatic scale is used to measure pitch.
	
	\item \textbf{Duration:} We use quarter length (ql) to measure the duration of a note. For example, a whole note is 4ql duration, and an eighth note is $\frac{1}{2}$ql duration.
\end{itemize}
For example, a note whose pitch is C1 and duration is 4ql duration can be represented as ``C14". Another example is shown in Figure \ref{fig:task}, the note sequence in this figure can be represented as a list of [a12, b12, c24, a12, e12], and the corresponding tagging sequence is [trills, none techniques, fermata, none techniques, mordent].

\section{Model}

The overview of the proposed model can be seen in Figure  \ref{fig:system}. Overall, this model is composed of three parts:

\begin{figure}[H]
	\centerline{
		\includegraphics[width=\columnwidth]{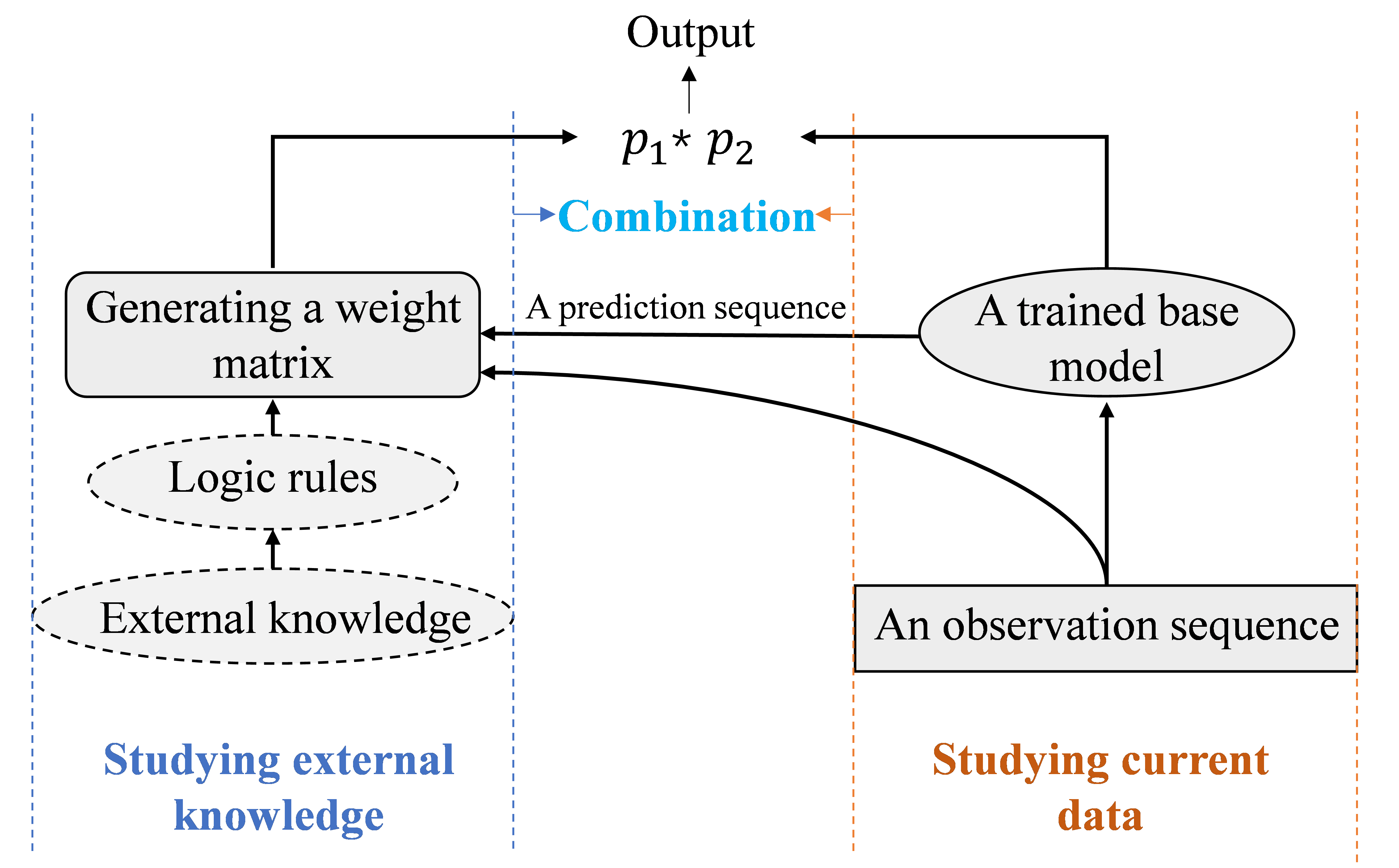}}
	{\caption{Model overview}  \label{fig:system} }
\end{figure}

\begin{itemize}
\item \textbf{Part 1: Studying current data} 

Model studies from current data using some general sequence tagging models (like CRF, BiLSTM, BiLSTM-CRF, and so on), then a trained base model can be gotten. Applying this trained model into an observation sequence, a prediction matrix $p_2$ (The number of columns represents the length of the sequence, the number of rows represents the number of the tag) and a prediction sequence can be gotten.  Besides, this prediction sequence and this observation sequence can be a part of the input of Part 2.

\item \textbf{Part 2: Studying external knowledge}

 In this paper, we focus on external knowledge that can be constructed into some logic rules. Using prediction sequence and observation sequence from Part 1, and some logic rules constructed from external knowledge. a weight matrix $p_1$ (has the same number of rows and columns as $p_2$) is finally generated (More details are in section 5.1). 

\item \textbf{Part 3: Combination}

 By calculating the Hadamard product of $p_1$ and $p_2$, the final output can be gotten. (More details are in section 5.2).
 
\end{itemize}
 
\subsection{Studying External Knowledge}

How to construct logic rules from external knowledge is different in different situations. Many methods have been discussed in discrete mathematics \cite{rosen2012discrete}. In this paper, we mainly describe how to use logic rules to generate a weight matrix based on state sequence (prediction sequence) and observation sequence.

At first, we should make out what kinds of logic rules are there in this playing techniques generation problem. Let $O$ be the observation sequence of length $T$, and $I$ be the corresponding state sequence. Then $I$ and $O$ can be represented as $I=\left\{i_1,i_2,... i_j...i_T \right\}$ and $O=\left\{o_1,o_2,... o_i...o_T \right\}$. Let $Tag$ be a tag set with $H$ elements, it can be represented as $Tag=\left\{tag_1,tag_2,... tag_k...tag_H \right\}$. Let $R$ be a login rule set. There are two kinds of logic rules in $R$:

\begin{itemize}
\item \textbf{Rule 1: Observation sequence constrains state sequence}

Let $F$ and $S$ represent predicates in logic rules. The statement $F(O)$ is the value of the propositional function $F$ at a observation sequence $O$. The statement $S(i_j, tag_k)$ is the value of the propositional function $S$ at a state and a tag. The corresponding logic rule can be represented as:

\begin{equation}\label{rule 1}
F(O) \Rightarrow \\
S(i_j, tag_k)
\end{equation}

The confidence of this logic rule is set as an adjustable parameter. We take a specific logic rule as an example, it is:

\begin{equation}\label{rule 2}
\begin{aligned}
duration(o_i) \textgreater 3 \Rightarrow
i_j = trills
\end{aligned}
\end{equation}

In this example, predicate $F$ refers to ``has a note whose duration is greater than 3",  predicate $S$ refers to ``The state corresponding with the note is".

\item \textbf{Rule 2: State sequence constrains state sequence}

Let $G$ and $H$ represent predicates in logic rules. The statement $G(I)$ is the value of the propositional function $F$ at a state sequence $I$. The statement $H(i_j, tag_k)$ is the value of the propositional function $H$ at a state and a tag. The corresponding logic rule can be represented as:

\begin{equation}\label{rule 3}
G(I) \Rightarrow \\
H(i_j, tag_k)
\end{equation}

The confidence of this logic rule is set as an adjustable parameter, too.

\end{itemize}

Then, a new weight matrix can be generated using Algorithm 1. Before this algorithm, the weight matrix $p_1$ (has the same size as the prediction matrix $p_2$) is initialized to a matrix with each element being 1. $h_1$ and $h_2$ is set as parameters to reflect the confidence of logic rules. 

\begin{table}[htbp]
	\begin{tabular*}{\columnwidth}{p{\columnwidth}l}
		\toprule
		\textbf{Algorithm 1:} Generating a new weight matrix
		\\ 
		\midrule
		\textbf{Input:} The old weight matrix $p_1$,\\
		$\qquad$ The observation sequence $O$,\\
		$\qquad$ The state sequence $I$,\\
		$\qquad$ The rule set $R$,\\
		$\qquad$ Parameters: $h_1$, $h_2$ $-$ Measure confidence \\
		1: \textbf{for} each rule in $R$: \\
		2: $\quad$ \textbf{if} $F(O)$ == True: \\
		3: $\qquad$ $P_1[j,k] \ *= h_1$  \\
		4: $\quad$ \textbf{if} $G(I)$ == True: \\
		5: $\qquad$ $P_1[j,k] \ *= h_2$  \\

		\textbf{Output:} A new weight matrix $p_1$,\\
		\bottomrule
	\end{tabular*}
\end{table}

\subsection{Combination}
\begin{figure}[H]
	\centerline{
		\includegraphics[width=\columnwidth]{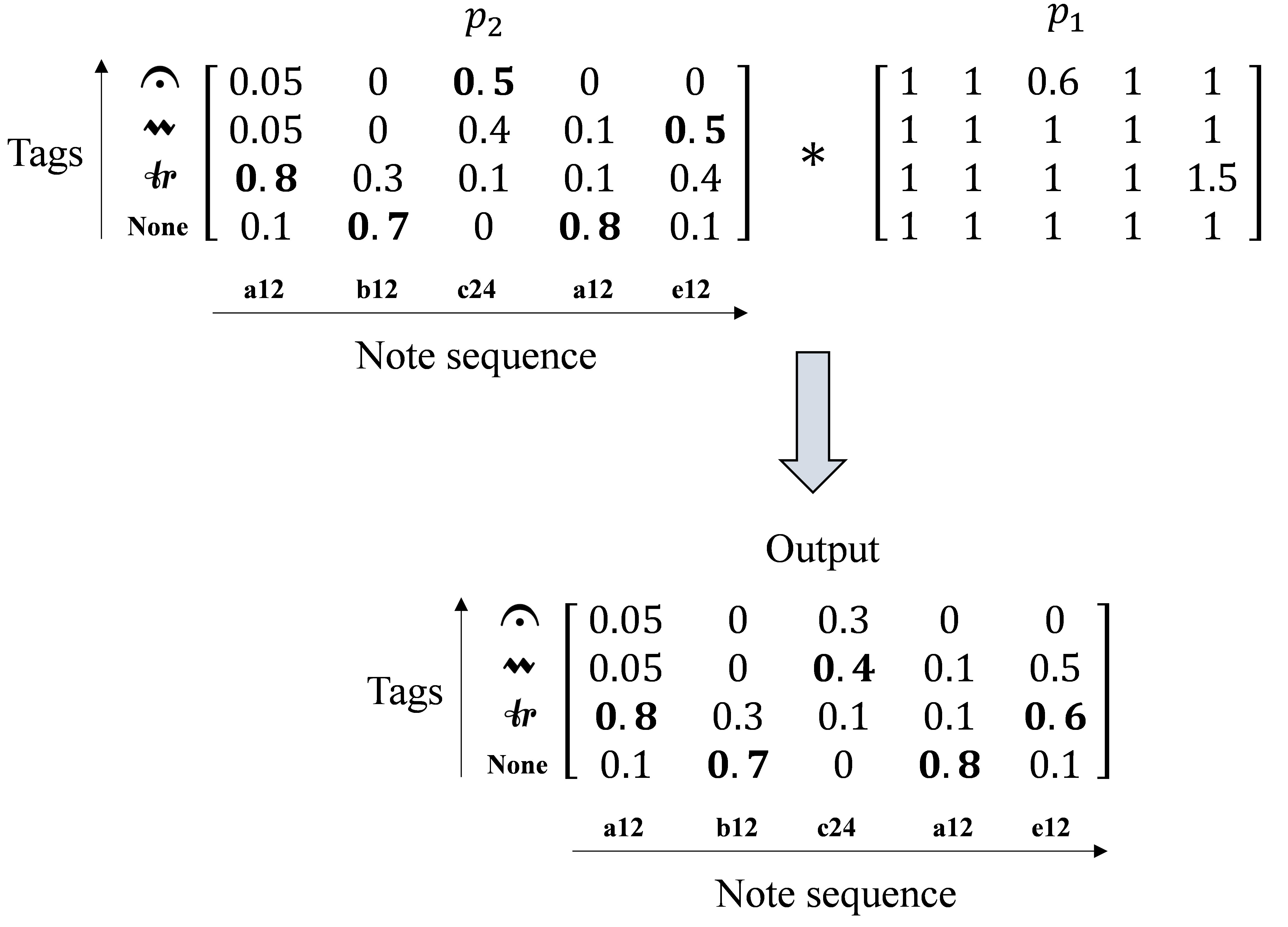}}
	{\caption{An example of combination}  \label{fig:p1p2} }
\end{figure}

By calculating the Hadamard product of $p_1$ and $p_2$, the final output can be gotten. An example is shown in Figure \ref{fig:p1p2}. Suppose this instrument has four kinds of tags. In this example, when external knowledge is not used, the prediction sequence is [trills, none techniques, fermata, none techniques, mordent]. When external knowledge is used, the prediction result is [trills, none techniques, mordent, none techniques, trills].

\section{Experiments}

\subsection{Setup}

Playing techniques can’t be generated without melodies. Therefore, our playing techniques generation experiment is based on the melodies that have been generated. We first used some music style transfer methods introduced in \cite{Zalkow2016} to generate melodies of a specific style, then used the model proposed in this paper to generate playing techniques.  We let humans evaluate the similarity between the generated music and the target music style, to compare the effect of generating only the melody with the effect of generating the melody and playing techniques. We use Chinese bamboo flute music as the experiment subject, and the dataset is from \cite{lizhen2003} and \cite{Dizi1993}. The data we used to train the playing techniques generation model includes 7320 notes in total. The data we used to test includes 4 pieces of melody.

The parameter setting of the proposed model in this paper is as follows. We use BiLSTM as the base model and use the external knowledge summarized in \cite{wanghe2014}, to construct 6 logic rules. We use a learning rate of 0.001. We set the dimension of the ``word" vector to 256, and the hidden layer size to 128. A batch size of 32 is used. We trained the model for 30 epochs. 

After generating symbolic music (melodies and playing techniques), we played them in Chinese bamboo flute to get audios, and used these audios to do the evaluation. Our evaluation was carried out with 35 participants. 12 of these participants have the experience of being a Chinese bamboo flute player, or have received formal education about Chinese bamboo flute music, or have work experience in this field. The other participants have other related music backgrounds. The evaluation score is in the scale from 1 to 10, where 1 represents the generated music is completely different from the target music style, 10 represents the generated music is very similar to the target music style.

\subsection{Results}
\begin{figure}[H]
	\centerline{
		\includegraphics[width=175pt]{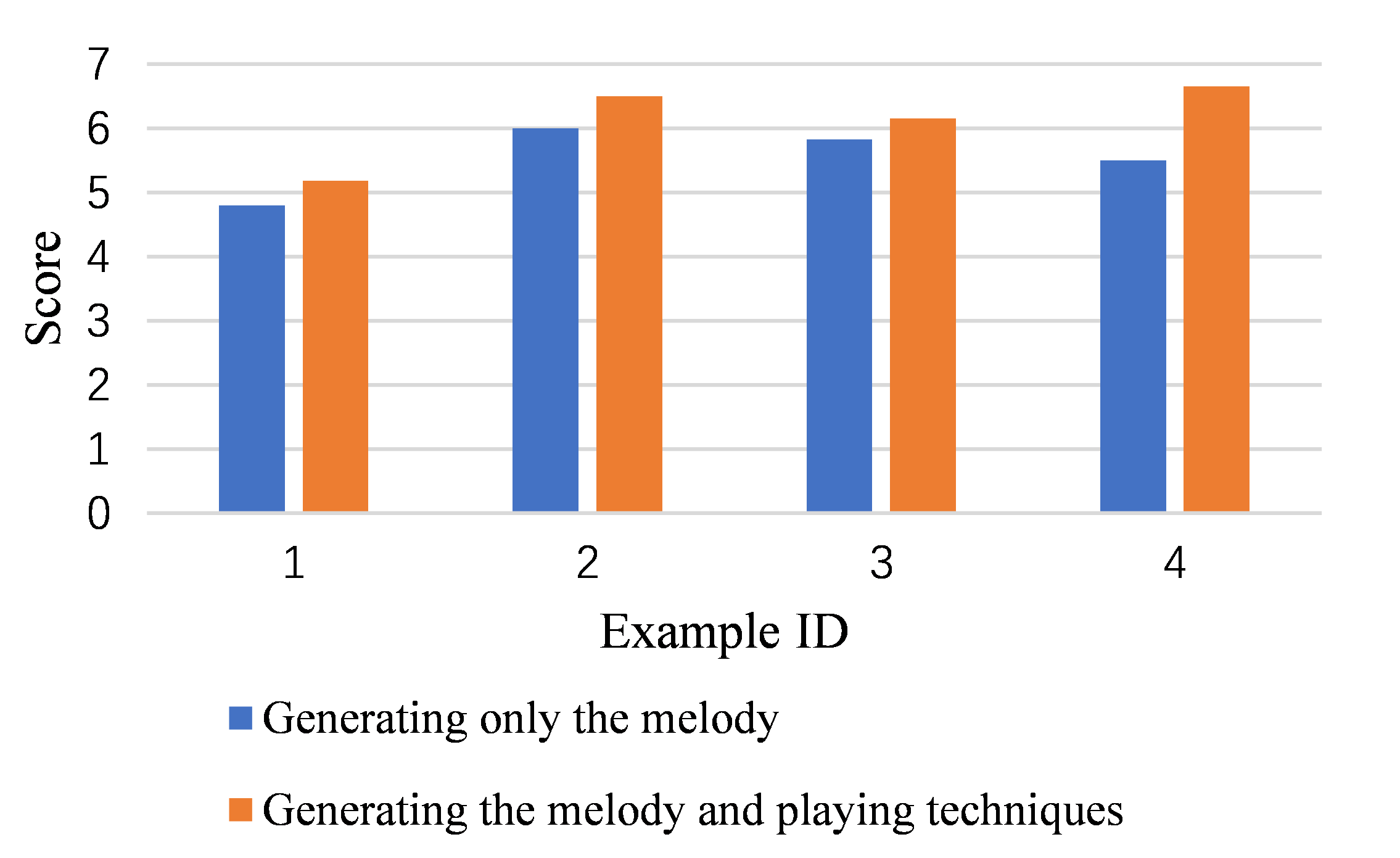}}
	{\caption{Experiment results}  \label{fig:exp} }
\end{figure}

The experiment results are shown in Figure \ref{fig:exp}. We can see that in all examples, the score is higher when the playing techniques generation model is used. This result shows that using our model can make music more lively compared with only generating melodies.

\subsection{A Generation Example}
\begin{figure}[H]
	\centerline{
		\includegraphics[width=\columnwidth]{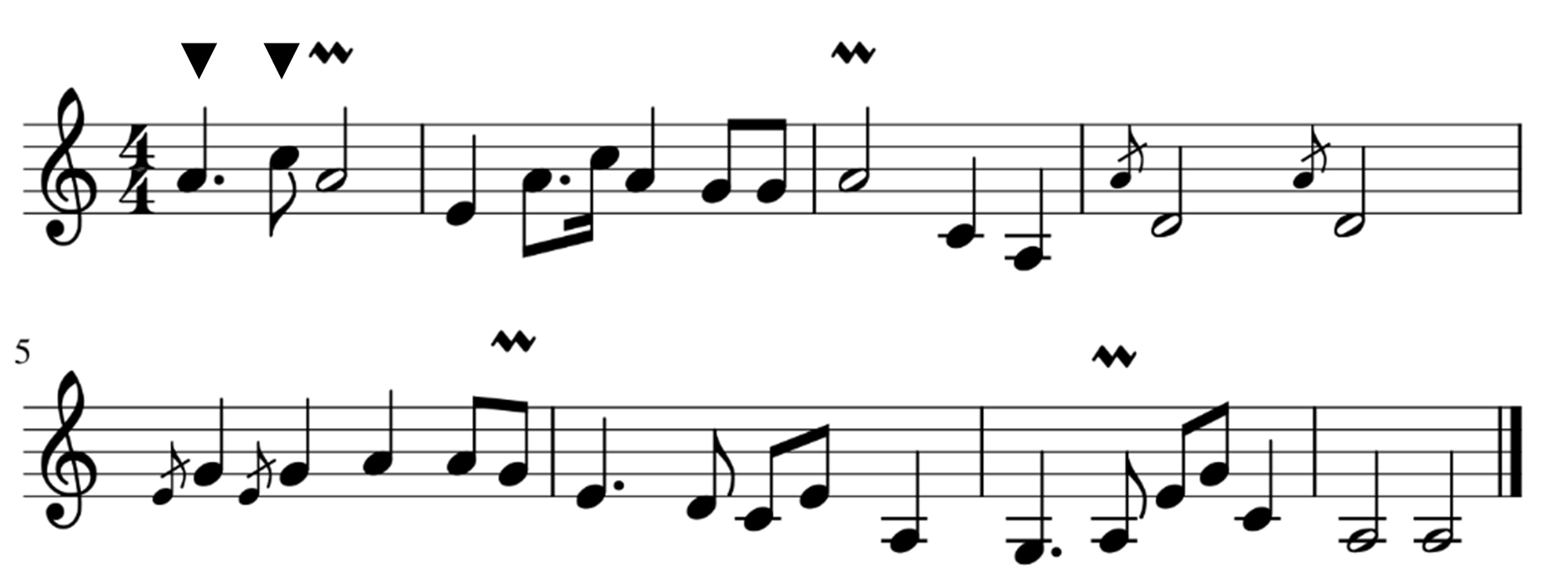}}
	{\caption{A generation example}  \label{fig:generation examples} }
\end{figure}

An example of generated music is shown in Figure \ref{fig:generation examples}. The generated music belongs to the style of the Northern school in Chinese bamboo flute. The generated playing techniques like tonguing and appoggiatura can make music style closer to the style of the Northern school. 

\section{Conclusion and Future Work}
Seeing playing techniques generation as a tagging problem, we have developed a framework that can use both the current data and external knowledge to generate playing techniques. Experiment results have shown that our proposed model can make generated music more lively.

There is still a lot of work to be done in the future, which includes more experiments and more applications. As a general playing techniques generation framework, it can be used not only in more music categories but also in other fields of music technology (e.g., music style transfer and music synthesis).

\bibliography{nlp4MusA}

\begin{thebibliography}{13}
\expandafter\ifx\csname natexlab\endcsname\relax\def\natexlab#1{#1}\fi

\bibitem[{Boulanger-Lewandowski et~al.(2012)Boulanger-Lewandowski, Bengio, and
  Vincent}]{10.5555/3042573.3042813}
Nicolas Boulanger-Lewandowski, Yoshua Bengio, and Pascal Vincent. 2012.
\newblock Modeling temporal dependencies in high-dimensional sequences:
  Application to polyphonic music generation and transcription.
\newblock In \emph{Proceedings of the 29th International Coference on
  International Conference on Machine Learning}, ICML’12, page 1881–1888,
  Madison, WI, USA. Omnipress.

\bibitem[{Dong et~al.(2017)Dong, Hsiao, Yang, and Yang}]{dong2017musegan}
Hao-Wen Dong, Wen-Yi Hsiao, Li-Chia Yang, and Yi-Hsuan Yang. 2017.
\newblock Musegan: Symbolic-domain music generation and accompaniment with
  multi-track sequential generative adversarial networks.
\newblock \emph{arXiv preprint arXiv:1709.06298}.

\bibitem[{Graves et~al.(2013)Graves, Jaitly, and Mohamed}]{graves2013hybrid}
Alex Graves, Navdeep Jaitly, and Abdel-rahman Mohamed. 2013.
\newblock Hybrid speech recognition with deep bidirectional lstm.
\newblock In \emph{2013 IEEE workshop on automatic speech recognition and
  understanding}, pages 273--278. IEEE.

\bibitem[{Hadjeres et~al.(2017)Hadjeres, Pachet, and
  Nielsen}]{hadjeres2017deepbach}
Ga{\"e}tan Hadjeres, Fran{\c{c}}ois Pachet, and Frank Nielsen. 2017.
\newblock Deepbach: a steerable model for bach chorales generation.
\newblock In \emph{International Conference on Machine Learning}, pages
  1362--1371.

\bibitem[{Hennig et~al.(2017)Hennig, Umakantha, and
  Williamson}]{hennig2017classifying}
Jay~A Hennig, Akash Umakantha, and Ryan~C Williamson. 2017.
\newblock A classifying variational autoencoder with application to polyphonic
  music generation.
\newblock \emph{arXiv preprint arXiv:1711.07050}.

\bibitem[{Huang et~al.(2015)Huang, Xu, and Yu}]{huang2015bidirectional}
Zhiheng Huang, Wei Xu, and Kai Yu. 2015.
\newblock Bidirectional lstm-crf models for sequence tagging.
\newblock \emph{arXiv preprint arXiv:1508.01991}.

\bibitem[{Lafferty et~al.(2001)Lafferty, McCallum, and
  Pereira}]{lafferty2001conditional}
John Lafferty, Andrew McCallum, and Fernando~CN Pereira. 2001.
\newblock Conditional random fields: Probabilistic models for segmenting and
  labeling sequence data.

\bibitem[{Li(2003)}]{lizhen2003}
Zhen Li. 2003.
\newblock \emph{Anthology of Zhen Li's Dizi music}.
\newblock People's Music Publishing House.

\bibitem[{Rosen and Krithivasan(2012)}]{rosen2012discrete}
Kenneth~H Rosen and Kamala Krithivasan. 2012.
\newblock \emph{Discrete mathematics and its applications: with combinatorics
  and graph theory}.
\newblock Tata McGraw-Hill Education.

\bibitem[{Wang(2014)}]{wanghe2014}
He~Wang. 2014.
\newblock Research on {C}hinese traditional bamboo flute playing techniques.
\newblock Master's thesis, Shaanxi Normal University.

\bibitem[{Yan and Yu(1994)}]{Dizi1993}
Niwen Yan and Yunfa Yu. 1994.
\newblock \emph{Collection of famous {C}hinese bamboo flute music}.
\newblock Shanghai Music Publishing House.

\bibitem[{Yang et~al.(2017)Yang, Chou, and Yang}]{yang2017midinet}
Li-Chia Yang, Szu-Yu Chou, and Yi-Hsuan Yang. 2017.
\newblock Midinet: A convolutional generative adversarial network for
  symbolic-domain music generation.
\newblock \emph{arXiv preprint arXiv:1703.10847}.

\bibitem[{Zalkow et~al.(2016)Zalkow, Brand, and Graf}]{Zalkow2016}
Frank Zalkow, Stephan Brand, and Bejamin Graf. 2016.
\newblock {M}usical style modification as an optimization problem.
\newblock In \emph{{P}roceedings of the {I}nternational {C}omputer {M}usic
  {C}onference}, pages 206--211.

\end{thebibliography}
\bibliographystyle{nlp4MusA_natbib}

\end{document}